\documentclass[conference]{IEEEtran}

\IEEEoverridecommandlockouts

\usepackage{url}
\usepackage{array}
\usepackage{cancel}
\usepackage{siunitx}
\usepackage{multirow}
\usepackage{graphicx}
\usepackage{textcomp}
\usepackage{multirow}
\usepackage{stfloats}
\usepackage{listings}
\usepackage{colortbl}
\usepackage{algorithm}
\usepackage[T1]{fontenc}
\usepackage[10pt]{moresize}
\usepackage[utf8]{inputenc}
\usepackage[noadjust]{cite}
\usepackage[noend]{algpseudocode}
\usepackage[table,x11names]{xcolor}
\usepackage{mathtools, cuted, nccmath}
\usepackage{amsmath,bm,amssymb,amsfonts}

\def\BibTeX{{\rm B\kern-.05em{\sc i\kern-.025em b}\kern-.08em
		T\kern-.1667em\lower.7ex\hbox{E}\kern-.125emX}
}

\graphicspath{{./figs/}}

\begin{document}
	\title{Multiplierless Design of High-Speed Very Large Constant Multiplications}
	
	\author{
	\IEEEauthorblockN{Levent~Aksoy\IEEEauthorrefmark{2},
			Debapriya~Basu~Roy\IEEEauthorrefmark{3},
			Malik~Imran\IEEEauthorrefmark{2}, 
			Samuel~Pagliarini\IEEEauthorrefmark{2}}
		\IEEEauthorblockA{\IEEEauthorrefmark{2}Department of Computer Systems, Tallinn University of Technology, Tallinn, Estonia\\
			Email: \{levent.aksoy, malik.imran, samuel.pagliarini\}@taltech.ee}
		\IEEEauthorblockA{\IEEEauthorrefmark{3}Computer Science and Engineering, IIT Kanpur, Kanpur, India\\
			Email: dbroy@cse.iitk.ac.in}}
	
	\maketitle
	
	\begin{abstract}
		In cryptographic algorithms, the constants to be multiplied by a variable can be very large due to security requirements. Thus, the hardware complexity of such algorithms heavily depends on the design architecture handling large constants. In this paper, we introduce an electronic design automation tool, called {\sc leiger}, which can automatically generate the realizations of very large constant multiplications for low-complexity and high-speed applications, targeting the ASIC design platform. {\sc leiger} can utilize the shift-adds architecture and use 3-input operations, i.e., \mbox{carry-save} adders (CSAs), where the number of CSAs is reduced using a prominent optimization algorithm. It can also generate constant multiplications under a hybrid design architecture, where 2-and 3-input operations are used at different stages. Moreover, it can describe constant multiplications under a design architecture using compressor trees. As a case study, \mbox{high-speed} Montgomery multiplication, which is a fundamental operation in cryptographic algorithms, is designed with its constant multiplication block realized under the proposed architectures. Experimental results indicate that {\sc leiger} enables a designer to explore the trade-off between area and delay of the very large constant and Montgomery multiplications and leads to designs with area-delay product, latency, and energy consumption values significantly better than those obtained by a recently proposed algorithm. 
	\end{abstract}
	
	
	\section{Introduction}

The Montgomery modular multiplication~\cite{montgomery85} is an essential operation in cryptographic algorithms, such as RSA~\cite{rsa78}, elliptic curve cryptography (ECC)~\cite{koblitz87}, and supersingular isogeny key encapsulation (SIKE)~\cite{jao11}. Since these algorithms operate on large prime numbers, e.g., 2048, 521, and 768 in RSA, ECC, and SIKE, respectively, the operands of the Montgomery multiplication are generally divided into smaller multiple bits, so that reasonable sizes of multiplication and addition operations can be used to compute the modular multiplication in acceptable latency~\cite{tenca99, huang11}. Note that the size of these multiple bits in the very large constant and input variable has a significant impact on the hardware complexity of the Montgomery multiplication design and thus, the exploration of values of these parameters is important to find the design, which fits perfectly in a \mbox{low-complexity} and high-speed application~\cite{huang11}.

The Montgomery multiplication includes the multiplication of a very large prime number by an input variable, called the \textit{very large constant multiplication} (VLCM) operation. There is only the algorithm of~\cite{aksoy22}, which aims to reduce the hardware complexity of the VLCM operation under the shift-adds architecture using only shift and \mbox{addition/subtraction} operations. To do so, it uses techniques, which maximize the sharing of common subexpressions among constant multiplications. However, it does not consider the high-speed realization of the VLCM operation, which is essential for high-performance cryptographic algorithms. To the best of our knowledge, there exist no algorithms proposed for the low-complexity and \mbox{high-speed} realization of the VLCM operation.

Thus, in this paper, we introduce an electronic design automation (EDA) tool, called {\sc leiger}, which can describe the high-speed design of the VLCM operation taking into account the area and targeting the ASIC design platform under three different architectures: (i)~the shift-adds architecture using carry-save adders (CSAs), denoted as SA-CSA; (ii)~the \mbox{shift-adds} architecture using 2-input adders/subtractors and \mbox{3-input} CSAs at different stages, denoted as SA-Hybrid; (iii)~the design architecture using compressor trees, denoted as CT. The very large constants are divided into smaller coefficients under the shift-adds architectures and the number of \mbox{2-input} operations and CSAs is reduced using optimization algorithms~\cite{hartley96, hosangadi1, spiral, hosangadi06}. The input variable is partitioned into smaller bits under the CT architecture and compressor trees are used to add the multiples of very large constants. Moreover, {\sc leiger} can automatically generate the entire Montgomery multiplication including its high-speed VLCM operation implemented under a given design architecture. Thus, the main contributions of this paper are two-fold: (i)~it proposes design architectures to realize the VLCM operation for high-speed applications, incorporating prominent algorithms to reduce its complexity; (ii)~it introduces high-speed Montgomery multiplication designs including the VLCM operation realized under the proposed architectures. It is observed from the experimental results that the exploration of the number of bits used in partitioning the very large constant and input variable in the VLCM operation is crucial while finding a \mbox{low-complexity} and high-speed design. It is shown that when compared to {\sc leiger}, the algorithm of~\cite{aksoy22} leads to VLCM operation and Montgomery multiplication designs with $4.3\times$ and $1.3\times$ larger area-delay product (ADP) values, respectively.

The rest of this paper is organized as follows: Section~\ref{sec:background} presents the background concepts. The proposed design architectures for the high-speed realization of the VLCM operation are described in Section~\ref{sec:architectures}. Experimental results are given in Section~\ref{sec:results} and finally, Section~\ref{sec:conclusions} concludes the paper.

	\section{Background}
\label{sec:background}

\subsection{Constant Multiplication}

The multiplication of constants by the variable $x$ can be written as the realization of constants by simply eliminating the variable. For example, $3x = x \ll 1 + x = (1 \ll 1 + 1)x$ can be written as $3 = 1 \ll 1$. These notations will be used interchangeably in this paper.

Since constants are determined beforehand and the realization of a multiplier in hardware is expensive in terms of area, the multiplication of constants by a variable is generally realized under the shift-adds architecture using only shift and addition/subtraction operations~\cite{nguyen}. Note that shifts can be realized using only wires, which represent no hardware cost. Thus, the optimization problem is to find the minimum number of adders/subtractors that are required to realize the constant multiplications. Note that this is an NP-complete problem~\cite{np-complete}. 

\begin{figure}[t]
	\centering
	\includegraphics[width=\linewidth]{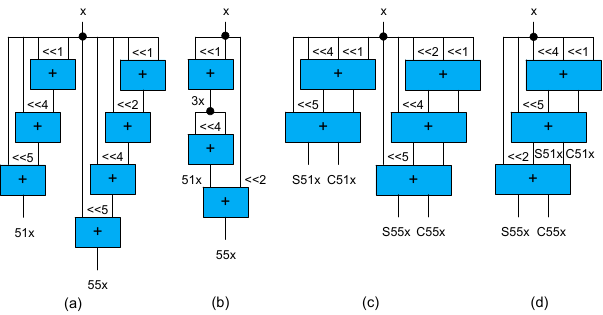}
	\vspace*{-8mm}
	\caption{Multiplierless design of $51x$ and $55x$: (a)~DBR technique~\cite{ercegovac} using 2-input operations; (b)~GB algorithm of~\cite{spiral}; (c)~DBR technique~\cite{ercegovac} using 3-input operations; (d)~CSE algorithm of~\cite{hosangadi06}.}
	\label{fig:cm}
	\vspace*{-6mm}
\end{figure}

In a straight-forward way, the digit-based recoding (DBR) technique~\cite{ercegovac} initially defines the constants under a number representation, e.g., binary, and then, for each nonzero digit in the representation of constant, it shifts the input variable based on the digit position and adds/subtracts the shifted variables according to the digit values. As an example, consider the constant multiplications $51x$ and $55x$. The decomposition of constants under the binary representation are given as follows: 
\begin{align}
	51x =& (110011)_{bin}x = x \ll 5 + x \ll 4 + x \ll 1 + x  \nonumber \\
	55x =& (110111)_{bin}x = x \ll 5 + x \ll 4 + x \ll 2 + x \ll 1 + x \nonumber
\end{align}
leading to a solution with 7 operations as shown in Fig.~\ref{fig:cm}(a).

% \begin{algorithm}[t]
	% 	\footnotesize
	% 	\caption{Montgomery Reduction}
	% 	\begin{algorithmic}[1]
		% 		\Statex \textbf{Input:} Odd modulus $M$ of $m$ bits, $R=2^m$, $gcd(R,M)=1$,  
		% 		\Statex $0<T<=RM$, precomputed $M'=-M^{-1}$ mod $R$.
		% 		\Statex \textbf{Output:} $t=T.R^{-1}$ mod $M$.
		% 		\State $Q=(T$ mod $R).M'$ mod $R$
		% 		\State $t=(T+Q.M)/R$
		% 		\If{$t>=M$}
		% 		\State $t=t-M$
		% 		\EndIf
		% 		\State \Return{t};
		% 	\end{algorithmic}
	% 	\label{algo:mont_reduc}
	% \end{algorithm}

The number of operations in a shift-adds design is generally reduced by sharing the partial products. To do so, many efficient common subexpression elimination (CSE ) and \mbox{graph-based} (GB) algorithms have been introduced. The CSE algorithms~\cite{hartley96,hosangadi1} initially define the constants under a number representation. Then, in an iterative fashion, they identify all possible subexpressions, which can be extracted from the nonzero digits in representations of constants, choose the "best" subexpression, generally the most common, and replace this subexpression with its realization. The GB algorithms~\cite{spiral, kumm-ilpmcm} are not restricted to any number representation and find the "best" intermediate constants, which enable to realize the constant multiplications with a small number of operations. For our example, the GB algorithm of~\cite{spiral} finds a solution with 3 operations as shown in Fig.~\ref{fig:cm}(b).

\begin{figure}[t]
	\centering
	\includegraphics[width=8.0cm]{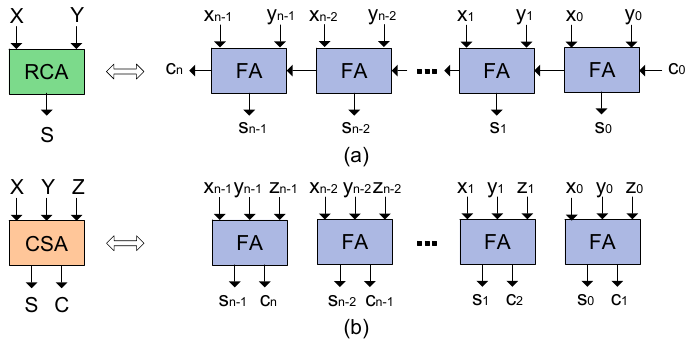}
	\vspace*{-4mm}
	\caption{Addition architectures: (a)~RCA; (b)~CSA.}
	\label{fig:csa}
	\vspace*{-6mm}
\end{figure}

In a shift-adds realization, an adder/subtractor is assumed to be a 2-input operation, which can be implemented using a \mbox{low-complexity} ripple carry adder (RCA) as shown in Fig.~\ref{fig:csa}(a). However, in high-speed applications, CSA is preferred to RCA~\cite{ienne}. CSA has three inputs and two outputs, i.e., sum (S) and carry (C), and an $n$-bit CSA includes $n$ full adders (FAs) as shown in Fig.~\ref{fig:csa}(b). Note that the delay of CSA is equal to the gate delay of an FA, independent of the input bit-width. The sum and carry outputs together form the computation, which can be obtained by adding these outputs using a fast adder at the end of the whole process. 

The DBR technique can find a shift-adds realization of constant multiplications using CSAs in a similar fashion. For our example, 5 CSAs are required as shown in Fig.~\ref{fig:cm}(c). Efficient CSE algorithms have also been proposed to reduce the number of CSAs~\cite{hosangadi06, sbcci08-I}. They iteratively determine all possible 3-term subexpressions and choose the "best" one to be shared among constant multiplications. For our example, the CSE algorithm of~\cite{hosangadi06} finds a solution with 3 CSAs as shown in Fig.~\ref{fig:cm}(d).

%Although a shift-adds realization using 2-input operations can be converted to a shift-adds design using CSAs~\cite{gustafsson04,ienne}, efficient CSE algorithms have been introduced to reduce the number of CSAs~\cite{hosangadi06, sbcci08-I}. 

\subsection{Montgomery Multiplication}
\label{subsec:montgomery}

Cryptographic algorithms like ECC are based on finite field arithmetic, which involves modular multiplication between large integers. Montgomery multiplication~\cite{montgomery85} allows us to perform modular reduction without computing any trial division. Algorithm~\ref{algo:mont_const} presents the constant time version of the Montgomery algorithm~\cite{orup1995simplifying}, where $M$ is the given prime modulus and $M'$ and $\overline{M}$ are precomputed constants. It performs \mbox{word-wise} multiplication between $a_i$ and $B$, where each $a_i$ is $r$ bits long, and involves the multiplication of the constant $\overline{M}$ by the variable $q_i$, which is the primary focus of this work.

%Given a prime modulus $M$, the Montgomery algorithm requires a precomputed constant  $M'$ such that $M'=-M^{-1} \ \text{mod} \ R$. The value of $R$ is chosen as $2^k>M$ with $k \in \mathcal{Z}^+$. Thus, the division and modular reduction by $R$ can be performed easily. 

\begin{figure}[t]
	\centering
	\vspace*{-2mm}
	\includegraphics[width=7.5cm]{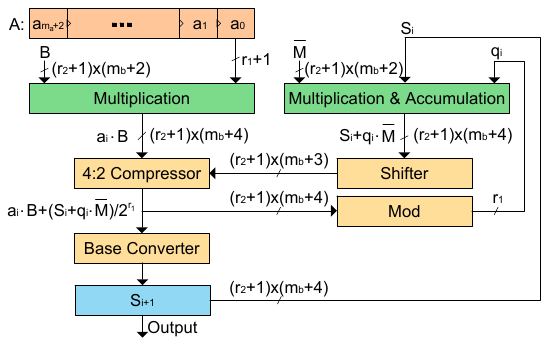}
	\vspace*{-4mm}
	\caption{RNS-based design of the Montgomery multiplication.}
	\label{fig:mont_arc}
	\vspace*{-6mm}
\end{figure}

There have been multiple works that focus on developing efficient architectures for the Montgomery multiplication, including systolic array-based architectures~\cite{mrabet2016systolic,ni2021high}. In this paper, we focus on the redundant number system \mbox{(RNS)-based} implementation of the Montgomery multiplication~\cite{roy19}. Note that RNS allows us to perform large integer arithmetic efficiently without long carry propagation. The architecture of the RNS-based Montgomery multiplication is shown in Fig.~\ref{fig:mont_arc}. The input $A$ is represented as a radix-$r_1$ redundant number with $1$ extra redundant bit for each word of dimension $r_1$. The input $B$ and the constant $\overline{M}$ are represented as \mbox{radix-$r_2$} redundant numbers. Note that $m_a$ and $m_b$ in Fig.~\ref{fig:mont_arc} are given as $\lceil m/r_1 \rceil$ and $\lceil m/r_2 \rceil$, respectively, where $m$ is the \mbox{bit-width} of the prime modulus $M$. In our ASIC design, the values of $r_1$ and $r_2$ are determined based on the parallel realization of multiplications. The \emph{Multiplication} and \emph{Multiplication \& Accumulation} blocks compute $a_i \cdot B$ and $S_i+q_i \cdot \overline{M}$ in a carry-save form, respectively. The output of the former block is combined with the output of the latter block after shifting via the \emph{4:2 compressor} circuit. The \emph{base converter} module converts the result of the carry-save form into \mbox{radix-$r_2$} redundant form and the \emph{mod} block realizes the modulo operation when the modulus is $2^{r_1}$.  

\begin{algorithm}[t]
	\footnotesize 
	\caption{Constant Time Montgomery Multiplication~\cite{orup1995simplifying}}\label{algo:mont_const}
	\begin{algorithmic}[1]
		\Statex \textbf{Input:} $M=\sum_{i=0}^{m-1}m_i\cdot 2^{ri}$, $A=\sum_{i=0}^{m+2}a_i \cdot 2^{ri}$ with $a_{m+2}=0$, 
		\Statex $B=\sum_{i=0}^{m+1}b_i\cdot 2^{ri}$, $M'=-M^{-1}$ mod $R$, $\overline{M}=(M'$ mod $2^r)\cdot M=\sum_{i=0}^{m+1}\overline{m}_i\cdot 2^{ri}$, $A,B<2\overline{M}$, $4\overline{M}<2^{rm}$, $R=2^{r(m+2)}$
		\Statex \textbf{Output:} $A$$\times$$B$$\times$$R^{-1} \ \text{mod} \ M$
		\State $S_0$=0
		\For{$i \gets 0$ \textbf{to} $m+2$}
		\State $q_i=S_i$ mod $2^r$
		\State $S_{i+1}$$=(S_i +q_i\cdot\overline{M})/2^r+a_i\cdot B$
		\EndFor
		\State \Return{$S_{m+3}=A \times B \times R^{-1}$ mod $M$ }
	\end{algorithmic}
\end{algorithm}

%In this paper, the implementation of the Montgomery multiplication on FPGAs using asymmetric unsigned multipliers in~\cite{roy19} is adapted to an ASIC implementation without a constraint on the multiplier size and its constant multiplication, i.e., $q_i \cdot \overline{M}$, is designed by our tool. Note that our tool can also be used in the implementation of three parallel modular multiplications described in~\cite{roy19}.

	\section{High-Speed Design of the VLCM Operation}
\label{sec:architectures}

In this section, we describe the realization of the VLCM operation under three design architectures, namely SA-CSA, SA-Hybrid, and CT, and introduce our EDA tool {\sc leiger}.

\subsection{SA-CSA Architecture} 

Given $n$ very large constants i.e., $lc_1, lc_2, \ldots, lc_n$, in hexadecimal format and the number of bits in partition, i.e., $p$, their shift-adds realization is obtained in three stages: (i)~partitioning; (ii)~realization of coefficients; and (iii)~realization of equations. These stages are described in the following sections.

\begin{figure}[t]
	\centering
	\includegraphics[width=9.0cm]{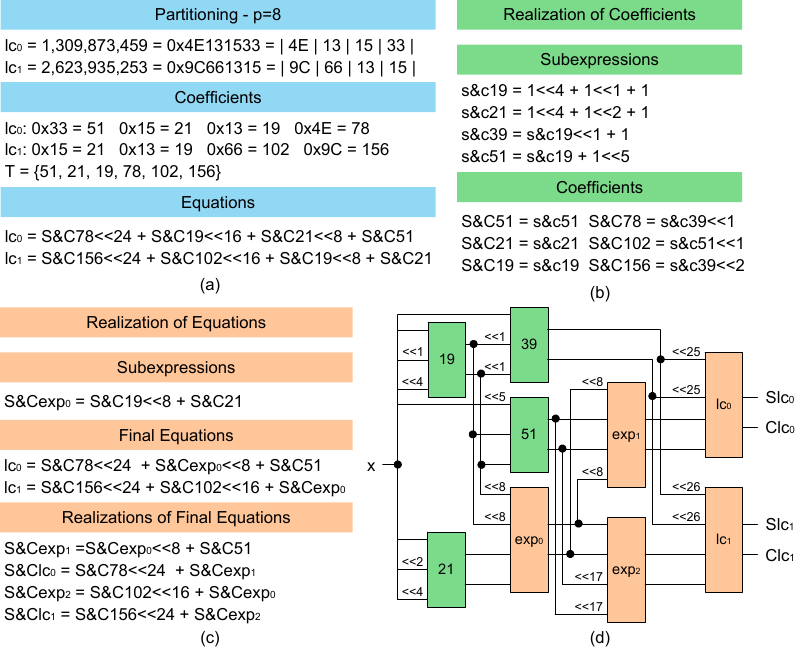}
	\vspace*{-7mm}
	\caption{Realization of the VLCM operation under the SA-CSA architecture: (a)~partitioning; (b)~coefficients; (c)~equations; (d)~implementation.}
	\label{fig:sa_csa}
	\vspace*{-6mm}
\end{figure}

\subsubsection{Partitioning}

Each very large constant $lc_i$, $1 \leq i \leq n$, is divided into $p$ bits, starting from the least significant bit and its $p$-bit coefficients, $c_1, c_2, \ldots, c_{d}$, where $d = \lceil w_i/p \rceil$ and $w_i$ is the bit-width of $lc_i$, are determined as $lc_i = \sum_{j=1}^{d} lc[jp-1 : (j-1)p] 2^{(j-1)p} = \sum_{j=1}^{d} c_j 2^{(j-1)p}$. The coefficients other than zero are stored as integers in a set called $T$ without repetition. Shift values of these coefficients are also computed based on the locations of these coefficients in $lc_i$ and stored in a set called $U$. Finally, the realization of each very large constant is written as an equation in the form of a summation of coefficients in set $T$ based on their shift values in set $U$, assuming that the multiplication of these coefficients by the input variable is realized using CSAs.

Fig.~\ref{fig:sa_csa} presents an example of the multiplierless realization of the VLCM operation. The partitioning step is given in Fig.~\ref{fig:sa_csa}(a) when $p$ is 8. Note that $S\&Ct_i$, $1 \leq i \leq |T|$ denotes the S and C outputs of CSA realizing $t_i$.

\subsubsection{Realization of Coefficients}

The CSE algorithm of~\cite{hosangadi06} is applied to find the shift-adds realization of coefficients in the set $T$ with a small number of CSAs. Fig.~\ref{fig:sa_csa}(b) presents its solution with 4 CSAs. 

\subsubsection{Realization of Equations}

Initially, the common subexpressions in equations are found using the CSE algorithm of~\cite{hosangadi06} and are replaced by their realizations. For our example, there is a single subexpression $exp_0$ as shown in Fig.~\ref{fig:sa_csa}(c). Since it has 4 inputs, 2 CSAs are required, implemented as $S\&Cexp_0 = C19 \ll 8 + Saux + Caux$ with $S\&Caux = S19 \ll 8 + C21 + S21$. Then, final equations are realized using CSAs considering the sizes of CSAs as shown in Fig.~\ref{fig:sa_csa}(c). For our example, each final equation requires 4 CSAs. 

Thus, the realization of the VLCM operation under the SA-CSA architecture requires a total of 14 CSAs, 4 for the coefficients, 2 for the common subexpression in equations, and 8 for the final equations as shown in Fig.~\ref{fig:sa_csa}(d). The realizations of equations are depicted using a 4-input operation, which actually includes 2 CSAs, for the sake of clarity in Fig.~\ref{fig:sa_csa}(d).

\subsection{SA-Hybrid Architecture}

To reduce the hardware complexity under the SA-CSA architecture, rather than CSAs, 2-input adders/subtractors can be used in the realization of coefficients at the second stage and in the realization of common subexpressions at the third stage, where the number of operations can be optimized using the GB algorithm of~\cite{spiral} and CSE algorithms of~\cite{hartley96, hosangadi1}, respectively. The final equations at the third stage can be realized using CSAs. Thus, the number of terms in the final equations, and consequently, the number of CSAs, can be reduced. 

\begin{figure}[t]
	\centering
	\includegraphics[width=8.5cm]{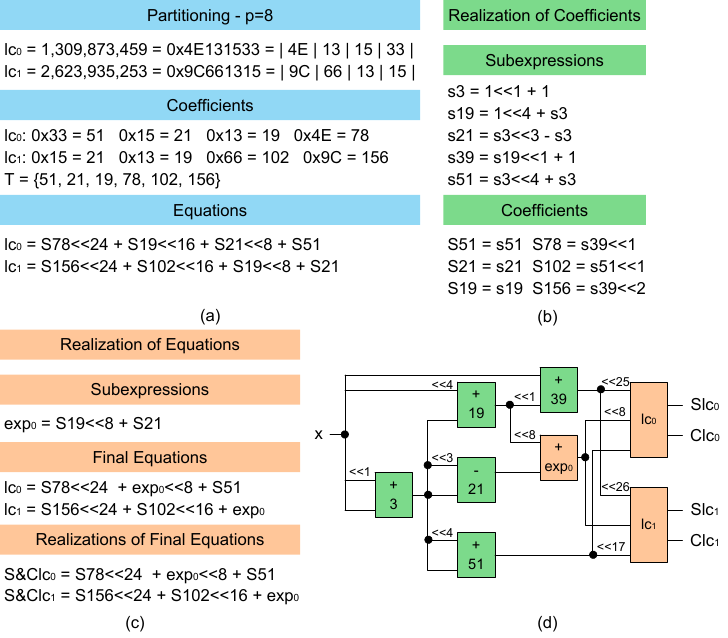}
	\vspace*{-4mm}
	\caption{Realization of the VLCM operation under the SA-Hybrid architecture: (a)~partitioning; (b)~coefficients; (c)~equations; (d)~implementation.}
	\label{fig:sa_hybrid}
	\vspace*{-4mm}
\end{figure}

Fig.~\ref{fig:sa_hybrid} presents the stages in the realization of the VLCM operation under the SA-Hybrid architecture for our example in Fig.~\ref{fig:sa_csa}. The realization of coefficients and the common subexpression requires 5 and 1 adders/subtractors, respectively. Also, each final equation requires 1 CSA. Thus, 6 2-input adders/subtractors and 2 CSAs are required as shown in Fig.~\ref{fig:sa_hybrid}(d). Note that the sign value shown in 2-input operations denotes the operation type in this figure. 

\begin{figure}[t]
	\centering
	\includegraphics[width=8.0cm]{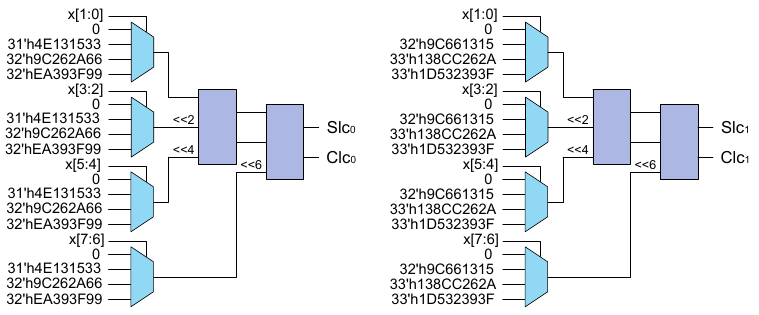}
	\vspace*{-4mm}
	\caption{Design of the VLCM operation in Fig.~\ref{fig:sa_csa} under the CT architecture.}
	\label{fig:ct_lcm_csa}
	\vspace*{-6mm}
\end{figure}

\subsection{CT Architecture}

For each very large constant $lc_i$, $1 \leq i \leq n$, the variable $x$ is partitioned into $r$ bits and the multiples of $lc_i$ between 0 and $2^r-1$, i.e., $0, lc_i, 2lc_i, \ldots, (2^r-1)lc_i$, are generated. For each partition of $x$, $x[jr-1:(j-1)r)]$, $1 \leq j \leq \lceil iw/r \rceil$, where $iw$ is the bit-width of the input variable $x$, the multiples of $lc_i$ are selected using MUXes, shifted accordingly, and added using compressor trees. In this case, $\lceil iw/r \rceil$ $2^r$-input MUXes and $\lceil iw/r \rceil - 2$ CSAs are required. The multiples of very large constants are described as constants so that the synthesis tool can apply its optimization techniques to simplify the logic.

For our example, the realization of constant multiplications is shown in Fig.~\ref{fig:ct_lcm_csa} when $iw$ and $r$ are 8 and 2, respectively.

\subsection{The EDA Tool}

Given the very large constants, the design architecture, the number of bits used in partitioning the constants or input variable, and other design parameters, {\sc leiger} automatically generates the behavioral description of the VLCM operation in Verilog, test-bench for verification, and synthesis and simulation scripts. It is equipped with algorithms developed for the optimization of the number of 2-input adders/subtractors and CSAs~\cite{spiral,hartley96,hosangadi1,hosangadi06}. It can generate the VLCM operation with one single output and two outputs as sum and carry. Moreover, it can automatically generate the Montgomery multiplication design described in Section~\ref{subsec:montgomery}, where the VLCM operation is realized under a given architecture. It is available at \textit{https://github.com/leventaksoy/vlcm}.

	\section{Experimental Results}
\label{sec:results}

As an experiment set, we use 5 elliptic curve instances taken from~\cite{safecurves}, namely \textit{anomalous}, \textit{anssifrp}, \textit{bn(2,254)}, \textit{brainpool256}, and \textit{brainpool348}, whose underlying primes do not have any special form and are 204, 256, 254, 256, and 384 bits long, respectively. In this section, we present the \mbox{gate-level} synthesis results of the VLCM block of the Montgomery multiplication and of the entire Montgomery multiplication design based on these elliptic curves. These designs are also implemented under the shift-adds architecture using 2-input operations~\cite{aksoy22}, denoted as SA-2IO. Note that logic synthesis was performed by Cadence Genus using a commercial 65\;nm cell library and designs were validated using 10,000 randomly generated inputs in simulation. 

\begin{figure}[t]
	\centering
	\vspace*{-4mm}
	\includegraphics[width=8.0cm]{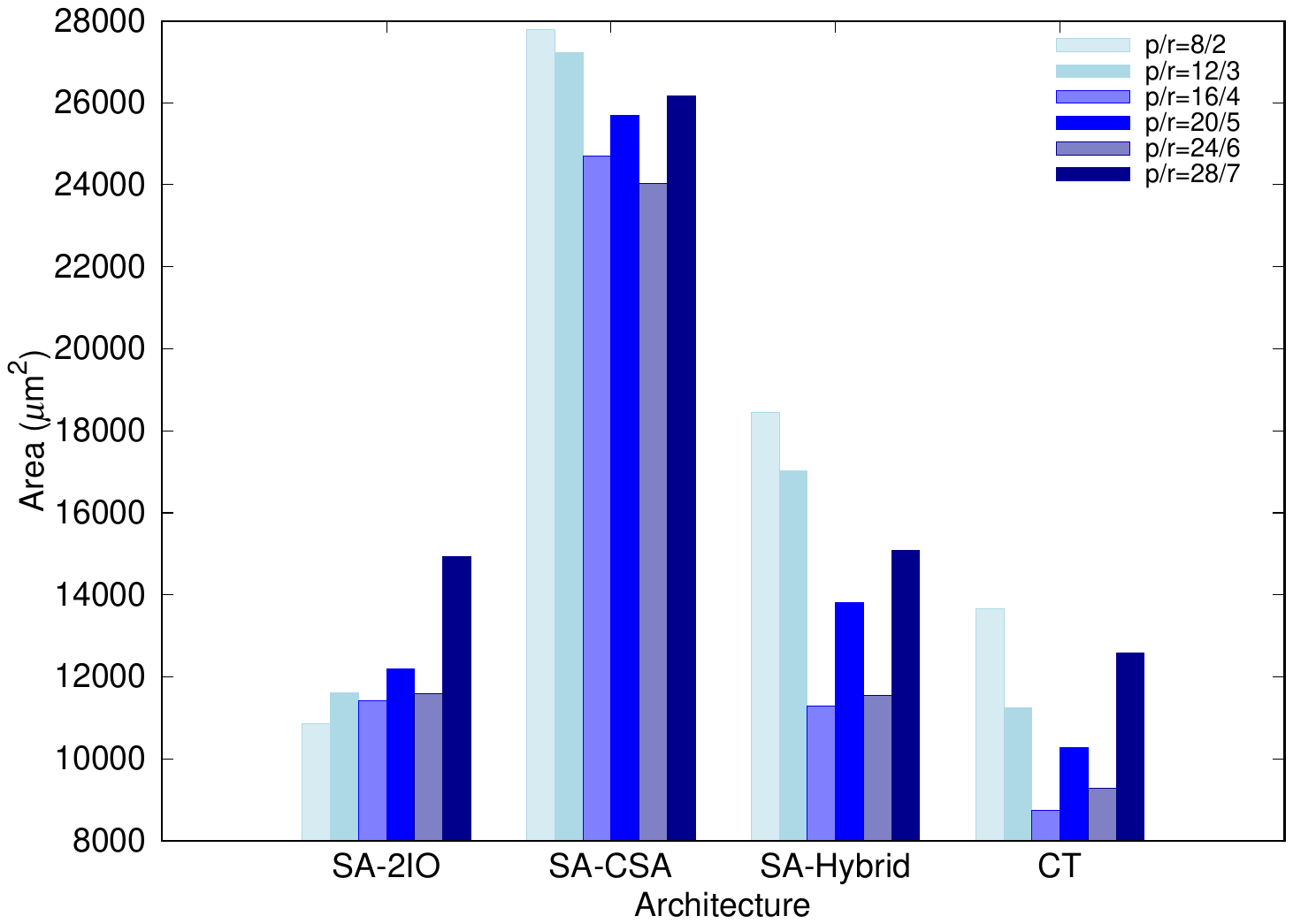}
	\vspace*{-8mm}
	\caption{Impact of $p$ and $r$ values on the area of the VLCM operation.}
	\label{fig:brainpool_conmul}
	\vspace*{-6mm}
\end{figure}

%\begin{figure*}[t]
%	\centering
%	\vspace*{-10mm}
%	\parbox{5.9cm}{\centerline{\includegraphics[width=6.8cm]{brainpool_area_conmul16}}}\
%	\parbox{5.9cm}{\centerline{\includegraphics[width=6.8cm]{brainpool_delay_conmul16}}}\
%	\parbox{5.9cm}{\centerline{\includegraphics[width=6.8cm]{brainpool_power_conmul16}}}\

%	\vspace*{-3.5mm}

%	\parbox{5.9cm}{\centerline{\scriptsize (a)}}\
%	\parbox{5.9cm}{\centerline{\scriptsize (b)}}\
%	\parbox{5.9cm}{\centerline{\scriptsize (c)}}\
%	\vspace*{-3mm}
%	\caption{Impact of $p$ and $r$ values on the hardware complexity of the VLCM operation: (a)~area; (b)~delay; (c)~power dissipation.}  
%	\label{fig:brainpool_conmul}
%\end{figure*}

\begin{figure*}[t]
	\centering
	\vspace*{-10mm}
	\parbox{5.9cm}{\centerline{\includegraphics[width=6.8cm]{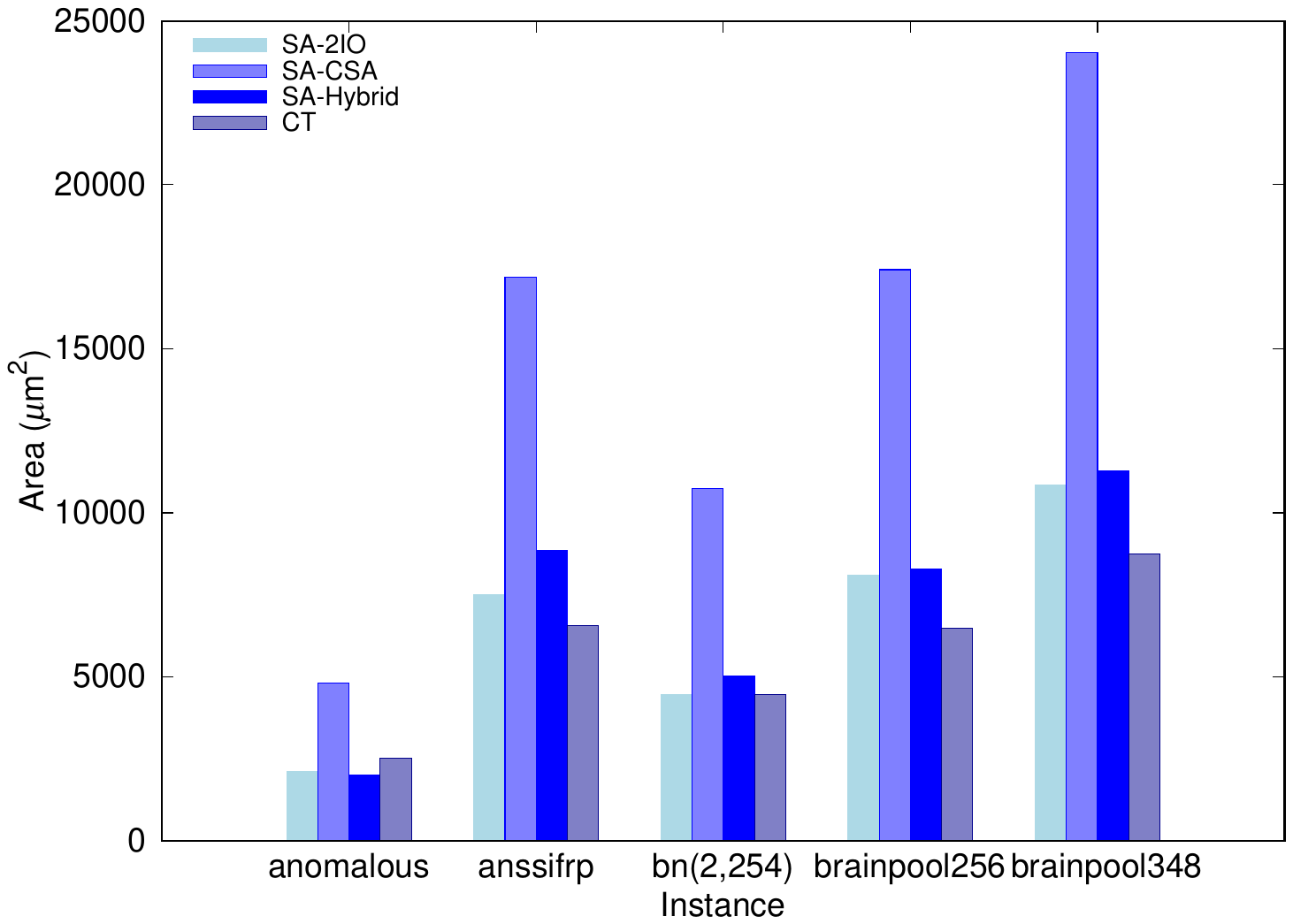}}}\
	\parbox{5.9cm}{\centerline{\includegraphics[width=6.8cm]{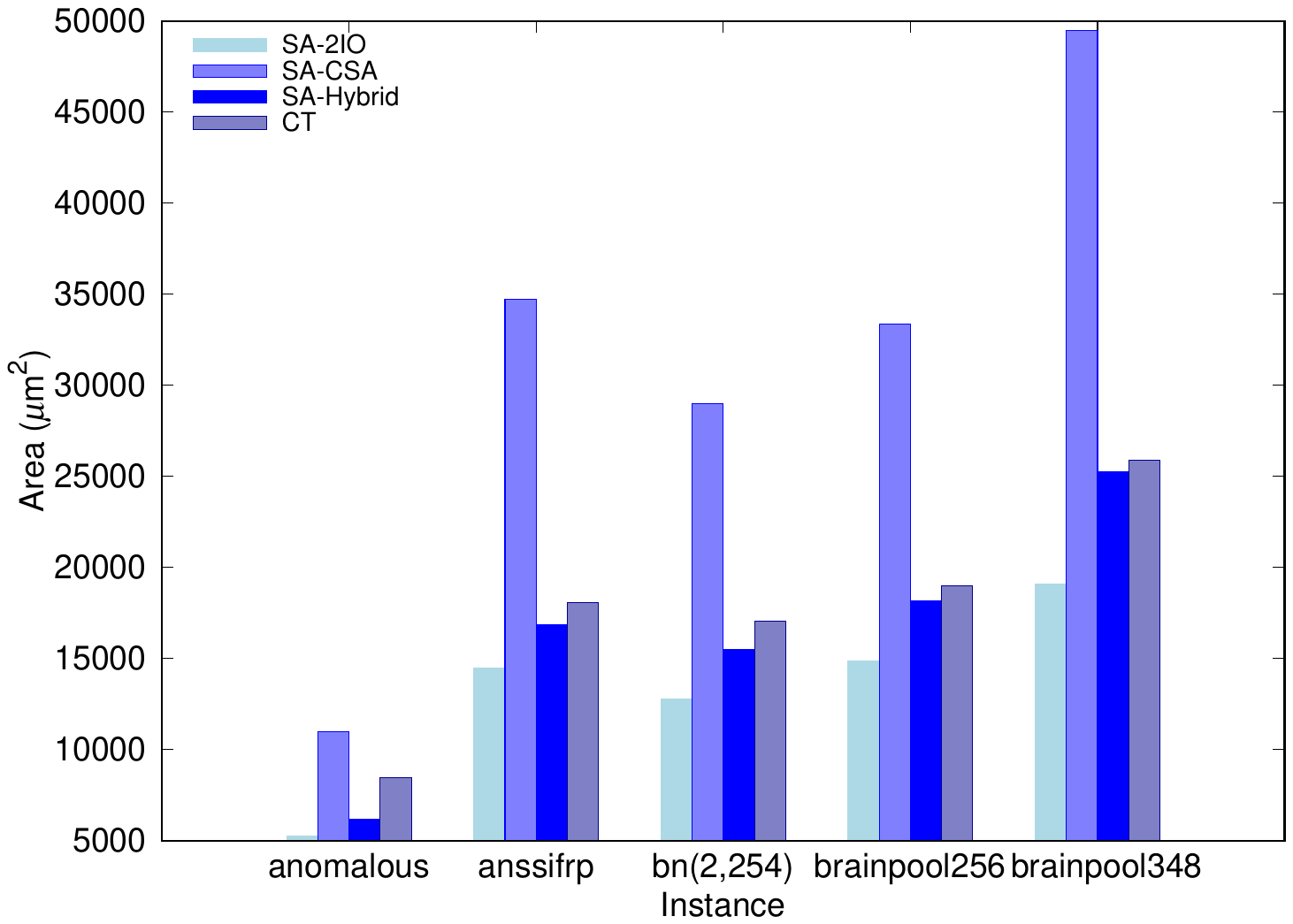}}}\
	\parbox{5.9cm}{\centerline{\includegraphics[width=6.8cm]{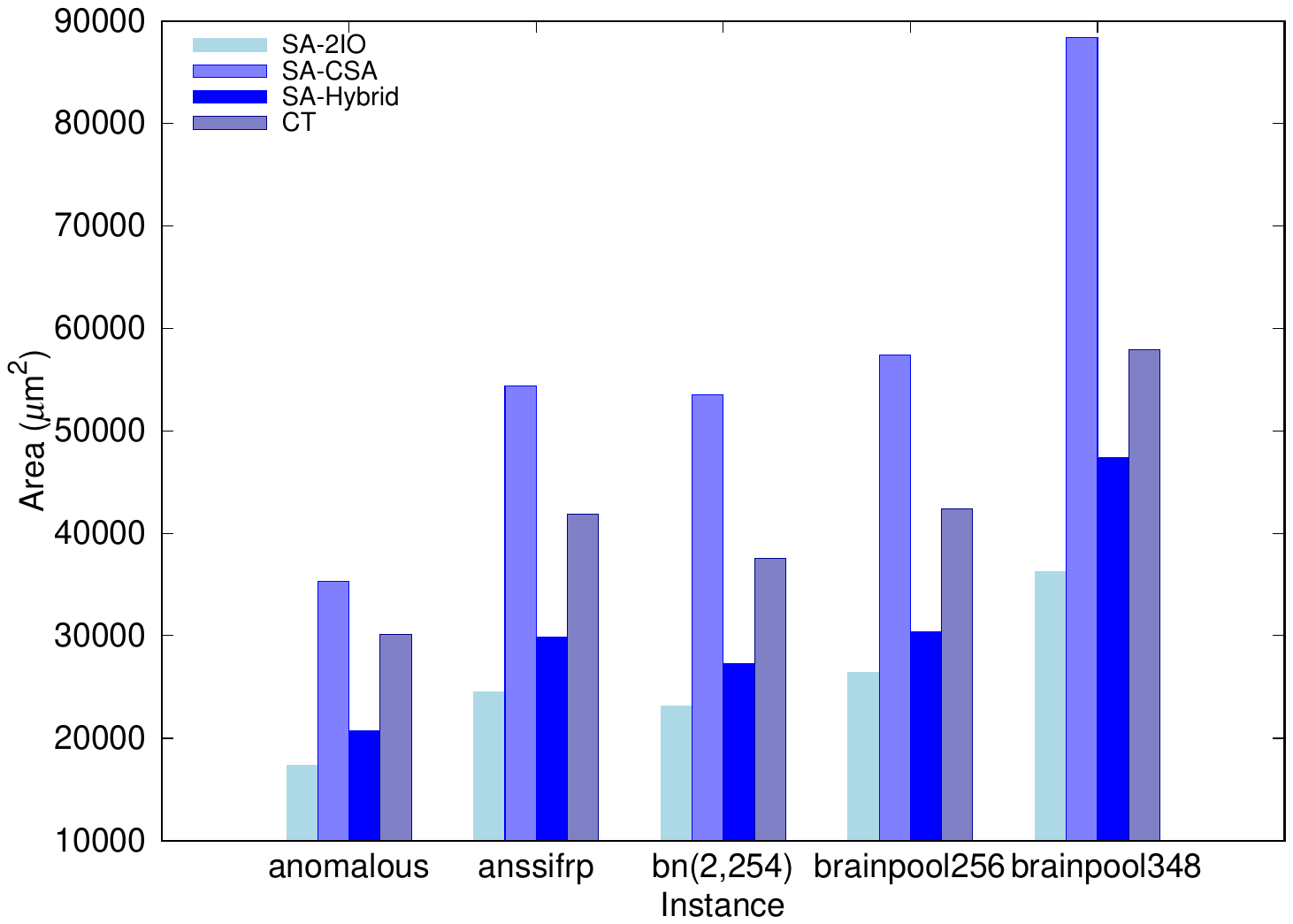}}}\
	
	\vspace*{-3.5mm}
	
	\parbox{5.9cm}{\centerline{\scriptsize (a)}}\
	\parbox{5.9cm}{\centerline{\scriptsize (b)}}\
	\parbox{5.9cm}{\centerline{\scriptsize (c)}}\
	\vspace*{-3mm}
	\caption{Impact of design architectures on the area of the VLCM operation: (a)~$iw$ is 16; (b)~$iw$ is 32; (c)~$iw$ is 64.}  
	\label{fig:conmul_area}
	\vspace*{-6mm}
\end{figure*}

\subsection{Very Large Constant Multiplication}
\label{subsec:cm}

As the first experiment, we generate the VLCM operations with two outputs as they are used in the Montgomery multiplication design. Possible realizations of the VLCM operation are obtained by changing the values of $p$, i.e., the number of bits in partitioning the constant, under the \mbox{shift-adds} architectures and $r$, i.e., the number of bits in partitioning the input variable, under the CT architecture. Note that $p$ ranges from 8 to 28 in a step of 4 and $r$ ranges between 2 and 7. The maximum value of $p$ is due to the limitation of the GB algorithm of~\cite{spiral} on the bit-width of constants. The designs are synthesized without a strict delay constraint aiming for area optimization. Fig.~\ref{fig:brainpool_conmul} shows the impact of $p$ and $r$ on the total area (in $\mu m^2$) of the \textit{brainpool348} instance under different architectures when the \mbox{bit-width} of the input variable, i.e., $iw$, is 16.

Observe from Fig.~\ref{fig:brainpool_conmul} that since the coefficients and equations to be realized under the shift-adds architectures are determined based on the $p$ value and the number of multiples of the very large constant, MUXes, and CSAs are determined based on the $r$ value under the CT architecture, they have a significant impact on area of the VLCM design. Note that the increase in area with respect to the minimum one can be up to 37.5\%, 56.3\%, 15.5\%, and 63.4\% under the SA-2IO, SA-CSA, \mbox{SA-Hybrid}, and CT architectures, respectively. Similar results were observed on other instances also when $iw$ is 32 and 64. 

To further explore the impact of a design architecture on the gate-level area of VLCM designs under different $iw$ values, Fig.~\ref{fig:conmul_area} presents the results belonging to VLCM designs with the smallest area value among others obtained with aforementioned $p$ and $r$ values when $iw$ is 16, 32, and 64.

Observe from Fig.~\ref{fig:conmul_area} that designs under the SA-CSA architecture have the largest area. When $iw$ is 16, the designs under the CT architecture have less area than designs under the \mbox{SA-2IO} and SA-Hybrid architectures, except the \textit{anomalous} instance. However, as $iw$ increases, the SA-2IO and \mbox{SA-Hybrid} architectures lead to designs with smaller area when compared to those realized under the CT architecture. For example, on the \textit{anssifrp} instance when $iw$ is 16, the design under the CT architecture has 12.3\% and 25.7\% gain in area with respect to the designs under the SA-2IO and SA-Hybrid architectures, respectively. However, on the same instance when $iw$ is 64, the design under the SA-2IO \mbox{(SA-Hybrid)} architecture has a 41.5\% (28.6\%) gain in the area when compared to the design under the CT architecture. This is because as $iw$ increases, both the size and number of CSA operations increase under the CT architectures while only the size of operations increases under the shift-adds architectures. 

%For example, on the \textit{brainpool348} instance when $iw$ is 16, the design under the SA-CSA architecture has $2.7\times$ area than the design under the CT architecture with the minimum area. 

\begin{table}[t]
	\centering
	\scriptsize
	\caption{VLCM designs with mad values.}
	\vspace{-3mm}
	\begin{tabular}{|@{\hskip3pt}l@{\hskip3pt}|c|c|@{\hskip3pt}c@{\hskip3pt}|c|c|c|@{\hskip3pt}c@{\hskip3pt}|c|}
		\hline
		\multirow{2}{*}{Architecture} & \multicolumn{4}{c|}{$iw$=16} & \multicolumn{4}{c|}{$iw$=32}\\ 
		\cline{2-9}
		& A & D & ADP & P & A & D & ADP & P\\
		\hline \hline		
		SA-2IO    & 5546  & 1341 & 7.4 & 1642 & 13847 & 1972 & 27.3 & 5601 \\
		SA-CSA    & 8978  & 738  & 6.6 & 2735 & 17711 & 942  & 16.6 & 6702 \\
		SA-Hybrid & 4873  & 1154 & 5.6 & 1410 & 13704 & 1777 & 24.3 & 5745 \\
		CT        & 5863  & 302  & 1.7 & 889  & 17684 & 505  & 8.9  & 4555 \\
		\hline
	\end{tabular}
	\label{tab:anomalous_conmul_minachdel}
	\vspace{-6mm}
\end{table}

To find the impact of the design architecture on the minimum achievable delay values, denoted as \textit{mad}, the \textit{anomalous} instance is synthesized under different architectures while the delay constraint is changed in a binary search manner until the minimum delay in the critical path is found without a negative slack. In this case, the initial lower and upper bounds on the delay constraint are set to 0\;ps and 80\;ns, respectively. Table~\ref{tab:anomalous_conmul_minachdel} presents the gate-level synthesis results of these designs when $iw$ is 16 and 32. In this table, \textit{A}, \textit{D}, \textit{ADP}, and \textit{P} are the total area in $\mu m^2$, the critical path delay in $ps$, area-delay product in $10^6 \times \mu m^2 \times ps$, and the total power dissipation in $\mu W$, respectively.

Observe from Table~\ref{tab:anomalous_conmul_minachdel} that while the SA-2IO architecture leads to designs with the largest \textit{mad} values, the designs under the CT architecture have the smallest \textit{mad}, ADP, and power dissipation values. The SA-Hybrid architecture achieves better \textit{mad} and ADP values than the SA-2IO architecture. The \mbox{SA-CSA} architecture leads to designs with promising \textit{mad} values, having the smallest ADP value among the shift-adds architectures when $iw$ is 32. This is because the logic synthesis tool has a large room to optimize area under a strict delay constraint when CSAs are used. We note that similar results were obtained on other elliptic curves also when $iw$ is 64.

\begin{figure}[t]
	\centering
	\vspace*{-4mm}
	\includegraphics[width=8.0cm]{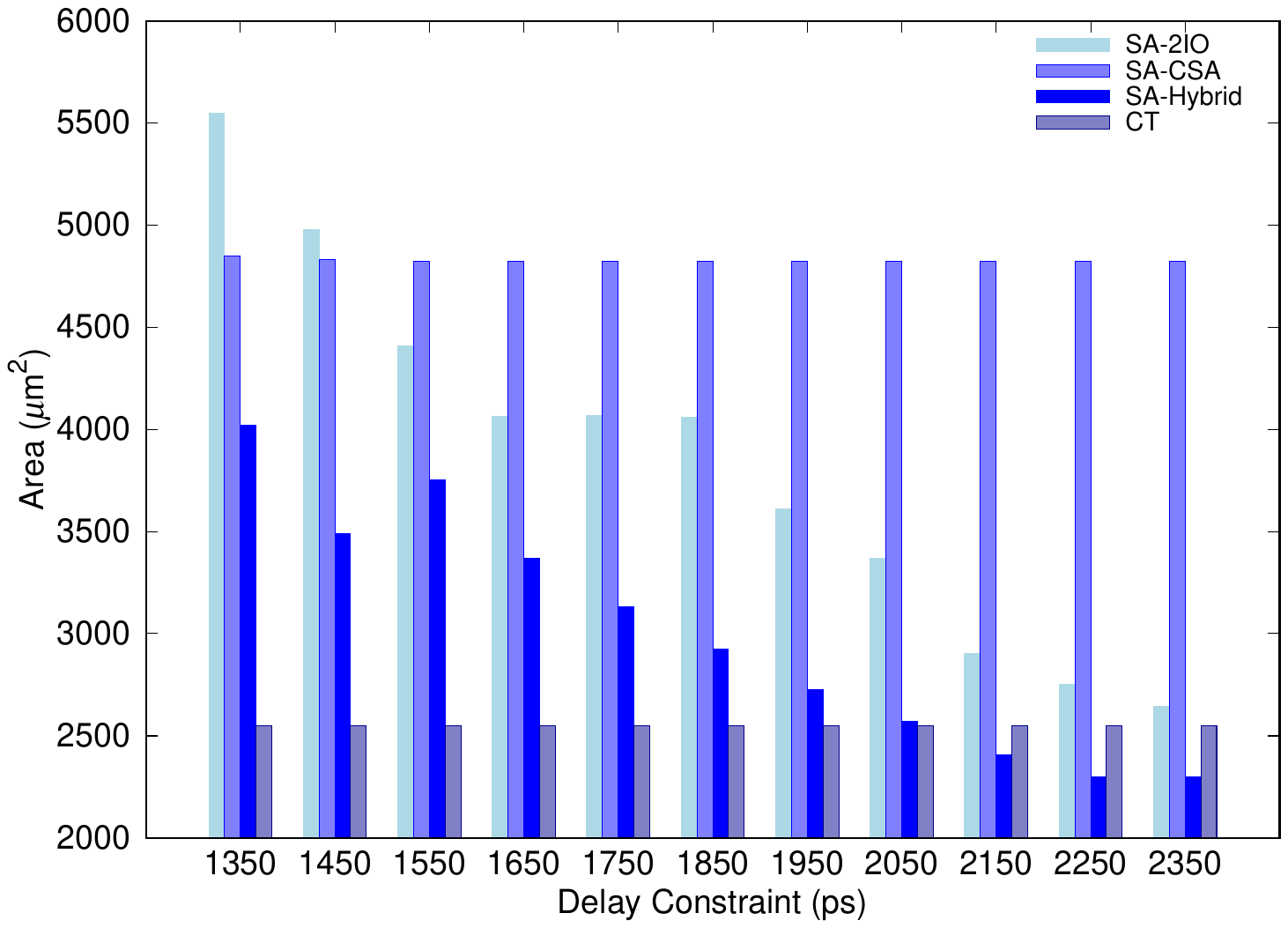}
	\vspace*{-9mm}
	\caption{Impact of delay constraint on the area of the VLCM operation.}
	\label{fig:anomalous_area_dc_16}
	\vspace*{-6mm}
\end{figure}

To further explore the area and delay tradeoff on the VLCM designs, the \textit{anomalous} instance is synthesized with a delay constraint ranging from 1350\;ps and 2350\;ps in a step of 100\;ps when $iw$ is 16. Observe from Table~\ref{tab:anomalous_conmul_minachdel} that the largest $mad$ value is 1341\;ps when $iw$ is 16. Fig.~\ref{fig:anomalous_area_dc_16} shows the \mbox{gate-level} area of these designs under different architectures.

Observe from Fig.~\ref{fig:anomalous_area_dc_16} that as the delay constraint is decreased, there is a slight increase in the area of designs under the \mbox{SA-CSA} and CT architectures. This is because the $mad$ values of these designs are smaller than the given delay constraints as shown in Table~\ref{tab:anomalous_conmul_minachdel} and hence, these delay constraints are easily satisfied by the logic synthesis tool. However, as the delay constraint is decreased, the area of designs under the \mbox{SA-2IO} and SA-Hybrid architectures increases significantly, which is simply to satisfy the delay constraint. Note that while the design under the SA-Hybrid architecture is 9.8\% smaller than that under the CT architecture when the delay constraint is 2350\;ps, the area of the design under the \mbox{SA-Hybrid} architecture is $1.5\times$ larger than that of the design under the CT architecture when the delay constraint is 1350\;ps.

\subsection{Montgomery Multiplication}
\label{subsec:mmm}

As the second experiment, we implement the entire Montgomery multiplication using the VLCM design realized under a given architecture. In these Montgomery multiplication designs, possible realizations of the VLCM operations were obtained by exploring the values of $p$ and $r$ as mentioned in Section~\ref{subsec:cm}. We note that different $p$ and $r$ values lead to Montgomery multiplication designs with different hardware complexity, although it is not as significant as in the VLCM operation shown in Fig.~\ref{fig:brainpool_conmul}. On the \textit{brainpool348} instance when $iw$ is 16, the increase in area with respect to the minimum one can be up to 2.8\%, 3.4\%, 2.5\%, and 5.2\% under the \mbox{SA-2IO}, SA-CSA, SA-Hybrid, and CT architectures, respectively. 

We also explored the impact of the design architecture of the VLCM operation on the gate-level area of the Montgomery multiplication design under different $iw$ values. Similar to the results shown in Fig.~\ref{fig:conmul_area}, the SA-CSA architecture in the VLCM operation leads to Montgomery multiplication designs with the largest area and as $iw$ increases, the SA-2IO and SA-Hybrid architectures lead to Montgomery multiplication designs with less area with respect to the CT architecture.

To explore the impact of the design architecture of the VLCM operation on the \textit{mad} value of the Montgomery multiplication design, the \textit{anomalous} instance is synthesized in a binary search manner as mentioned in Section~\ref{subsec:cm}. Table~\ref{tab:anomalous_monmul_minachdel} presents the gate-level synthesis results of these designs. In this table, \textit{L} denotes the latency of the design in $ns$, computed as $D \times CC$, where $CC$ is the number of clock cycles required to compute the multiplication result and is 51 and 30 for the \textit{anomalous} instance when $iw$ is 16 and 32, respectively. Also, \textit{E} denotes the energy consumption in $nW$, computed as $L \times P$.

Observe from Table~\ref{tab:anomalous_monmul_minachdel} that the CT architecture leads to designs with the smallest \textit{mad}, ADP, latency, and energy consumption values, but with the largest area. The designs under the \mbox{SA-CSA} and \mbox{SA-Hybrid} architectures have smaller \textit{mad}, ADP, latency, and energy consumption values when compared to those under the \mbox{SA-2IO} architecture. 

\begin{table}[t]
	\centering
	\scriptsize
	\caption{Montgomery multiplication designs with mad values.}
	\vspace{-3mm}
	\begin{tabular}{|@{\hskip2pt}l@{\hskip2pt}|c@{\hskip2pt}|@{\hskip2pt}c@{\hskip2pt}|@{\hskip2pt}c@{\hskip2pt}|@{\hskip2pt}c@{\hskip2pt}|@{\hskip2pt}c@{\hskip2pt}|c@{\hskip2pt}|@{\hskip2pt}c@{\hskip2pt}|@{\hskip2pt}c@{\hskip2pt}|@{\hskip2pt}c@{\hskip2pt}|@{\hskip2pt}c@{\hskip2pt}|}
		\hline
		\multirow{2}{*}{Architecture} & \multicolumn{5}{c|}{$iw$=16} & \multicolumn{5}{c|}{$iw$=32}\\ 
		\cline{2-11}
		& A & D & ADP & L & E & A & D & ADP & L & E \\
		\hline \hline		
		SA-2IO    & 74220  & 1680 & 124.6 & 85.6 & 10683 & 117214 & 2166 & 253.8 & 64.9 & 16497 \\
		SA-CSA    & 94632  & 1087 & 102.8 & 55.4 & 5703  & 197744 & 1197 & 236.6 & 35.9 & 8500  \\
		SA-Hybrid & 73685  & 1550 & 114.2 & 79.0 & 9028  & 123590 & 1914 & 236.5 & 57.4 & 13583 \\
		CT        & 106287 & 875  & 93.0  & 44.6 & 4150  & 174710 & 1158 & 202.1 & 34.7 & 7028  \\
		\hline
	\end{tabular}
	\label{tab:anomalous_monmul_minachdel}
	\vspace{-6mm}
\end{table}

Finally, Table~\ref{tab:monmul_minachdel} presents the high-speed Montgomery multiplication designs under the CT architecture with the \textit{mad} values when $iw$ is 16 and 32.

Observe from Table~\ref{tab:monmul_minachdel} that as $iw$ increases, the area and energy consumption of the Montgomery multiplication design increase significantly. However, in this case, the number of clock cycles and latency decrease. Note that while the gain in area can be up to 42.2\% on the \textit{brainpool348} instance when compared to designs generated when $iw$ is 16 and 32, the gain in latency can be up to 28.2\% on the \textit{anssifrp} instance when compared to designs generated when $iw$ is 32 and 16.

%Note that the use of \mbox{high-speed} Montgomery multiplication designs has a significant impact on the latency of the entire cryptographic algorithm~\cite{ding19}.

	\section{Conclusions}
\label{sec:conclusions}

This paper introduced an EDA tool, called {\sc leiger}, that can generate high-speed realizations of the VLCM operation. {\sc leiger} can implement the VLCM operation under different architectures and is equipped with techniques, which can optimize the number of operations used in the \mbox{shift-adds} architectures. As a case study, {\sc leiger} was applied to the VLCM block of the Montgomery multiplication and \mbox{high-speed} Montgomery multiplication designs were obtained. It was shown that {\sc leiger} can enable a designer to generate realizations of both VLCM and Montgomery multiplication designs, which can fit into a low-complexity and high-speed application, and to explore the tradeoff between area and delay of the VLCM and Montgomery multiplication designs. 

\begin{table}[t]
	\centering
	\scriptsize
	\caption{High-speed Montgomery multiplication designs.}
	\vspace{-3mm}
	\begin{tabular}{|@{\hskip3pt}l@{\hskip3pt}|c@{\hskip3pt}|@{\hskip3pt}c@{\hskip3pt}|@{\hskip3pt}c@{\hskip3pt}|@{\hskip3pt}c@{\hskip3pt}|c@{\hskip3pt}|@{\hskip3pt}c@{\hskip3pt}|@{\hskip3pt}c@{\hskip3pt}|@{\hskip3pt}c@{\hskip3pt}|}
		\hline
		\multirow{2}{*}{Instance} & \multicolumn{4}{c|}{$iw$=16} & \multicolumn{4}{c|}{$iw$=32}\\ 
		\cline{2-9}
		& A & CC & L & E & A & CC & L & E \\
		\hline \hline		
		anomalous    & 106287 & 51 & 44.6 & 4150  & 174710 & 30 & 34.7 & 7028 \\
		anssifrp     & 133381 & 63 & 60.1 & 7664  & 224574 & 36 & 43.1 & 11623\\
		bn(2,254)    & 133850 & 60 & 56.6 & 7157  & 211212 & 36 & 41.0 & 9864 \\
		brainpool256 & 140497 & 63 & 58.9 & 7738  & 230367 & 36 & 42.3 & 11469\\
		brainpool348 & 178034 & 87 & 79.5 & 12939 & 308665 & 48 & 58.2 & 21800\\
		\hline
	\end{tabular}
	\label{tab:monmul_minachdel}
	\vspace{-6mm}
\end{table}
	
	\section*{Acknowledgment}

This work was partially supported by the EU through the European Social Fund in the context of the project “ICT programme”. This work was also initiated as part of the EU’s H2020 project SAFEST (grant agreement No 952252).
	
	\bibliographystyle{IEEEtran}
	\bibliography{aspdac24}
\end{document}